\documentclass[usegraphicx,usenatbib]{mn2e}
\input psfig

\begin{document}
\arraycolsep0.35mm                      
\def\ex#1{\langle#1\rangle}
\def\d{{\rm d}}\newcommand{\e}{{\rm e}}\def\b#1{{\bf#1}}
\def\e{{\rm e}}
\def\i{\relax\ifmmode{\rm i}\else\char16\fi}
\def\fracj#1#2{{\textstyle{#1\over#2}}}
\newcommand{\be}{\begin{equation}}
\newcommand{\ee}{\end{equation}}
\newcommand{\Gyr}{\,{\rm Gyr}}
\newcommand{\pc}{\,{\rm pc}}
\newcommand{\pa}{\partial}
\newcommand{\kpc}{\,{\rm kpc}}
\newcommand{\kms}{\,{\rm km}~{\rm s}^{-1} }
\newcommand{\sm}{\,{\rm M}_{\sun}}
\newcommand{\mjup}{\,M_{\rm J}}
\newcommand{\au}{\,{\rm AU}}
\newcommand{\den}{\,{\rm M}_{\sun}\mbox{pc}^{-3}}
\newcommand{\dc}{D_{\rm c}}
\newcommand{\nc}{n_{\rm c}}
\newcommand{\ns}{n_{\rm s}}
\newcommand{\ds}{D_{\rm s}}
\newcommand{\vl}{v_{\rm l}}
\newcommand{\vs}{v_{\rm s}}
\newcommand{\ovs}{\overline{\b v}_{\rm s}}
\newcommand{\vo}{v_{\rm o}}
\newcommand{\vrot}{v_{\rm rot}}
\newcommand{\Gz}{\Gamma_0}


\title[Radial mixing in galactic discs]
{Radial mixing in galactic discs}
\author[Sellwood \& Binney]{J. A. Sellwood$^1$ \& J. J. Binney$^2$\\
$^1$Department of Physics and Astronomy, Rutgers University, 
 136 Frelinghuysen Road, Piscataway, NJ 08854, USA\\
$^2$Theoretical Physics, Department of Physics, University of Oxford,
 Keble Rd, Oxford OX1 3NP.}
\maketitle

\begin{abstract}
We show that spiral waves in galaxy discs churn the stars and gas in a manner that largely preserves the overall angular momentum distribution and leads to little increase in random motion.  Changes in the angular momenta of individual stars are typically as large as $\sim50\%$ over the lifetime of the disc.  The changes are concentrated around the corotation radius for an individual spiral wave, but since transient waves with a wide range of pattern speeds develop in rapid succession, the entire disc is affected.  This behaviour has profound consequences for the metallicity gradients with radius in both stars and gas, since the ISM is also stirred by the same mechanism.  We find observational support for stirring, propose a simple model for the distribution of stars over metallicity and age, and discuss other possible consequences.
\end{abstract}


\section{Introduction} \label{intro}

The metallicity of the ISM increases through the life of a galaxy as dying stars throw newly formed heavy elements into the interstellar medium (ISM).  It is well-established that the metallicity of the ISM decreases outwards in the Milky Way (Wilson \& Rood 1994) and other galaxies (Henry \& Worthey 1999).  Since stars have the metallicity of the material from which they were formed, we should expect: (i) older stars to be more metal-poor than younger ones, and (ii) for a given age, the metallicity of a star to be a decreasing function of the radius at which it was born.

If there were no stellar migration, the stars at a given radius would have metallicities that varied from zero up to the present-day value in the ISM at that radius and there would be a perfect correlation between the ages and metallicities of stars.  Edvardsson et al.\ (1993) do not find a tight correlation between ages and metallicities of stars in the solar neighbourhood, but find instead that stars of a given age have a broad spread in metallicity; e.g.\ stars of ages $\sim 5\;$Gyr have $-0.7 \la \hbox{[Fe/H]} \la 0.2$.  Only the lower bound of the metallicity distribution decreases with age -- there are no young, metal-poor stars in the solar neighbourhood.

Because of a general increase of non-circular motion with age (Wielen 1977; Lacey 1984; Edvardsson et al.; Dehnen \& Binney 1998), older stars oscillate more in radius than do younger ones, causing stars from an increasing spread of `home radii' to appear in a sample at any radius.  Typical radial excursions for a population of stars with rms radial velocity $\sigma_{\rm u}$ are $\Delta \simeq \sqrt{2}\sigma_{\rm u}/\kappa$, where is $\kappa$ is the epicycle frequency.  For old stars in the thin disc near the Sun $\Delta\sim 1.3\kpc$, and migration over distances of this order proves insufficient to explain the weakness of the correlation between age and metallicity found by Edvardsson et al.

The spread in the age-metallicity relation therefore requires either that the metallicity of the ISM was much less homogeneous in the past than it is today, a distinctly unattractive possibility, or that stars have migrated by more than their epicycle size from their radii of birth, as proposed by Wielen, Fuchs \& Dettbarn (1996).  Since the home radius is determined by the star's specific angular momentum, $L$, about the Galaxy's symmetry axis, the latter possibility requires $L$ to be changed.

Changes in angular momentum can arise only from non-axisymmetric forces, and two sources have been identified: molecular clouds (Spitzer \& Schwarzschild 1953) and spiral arms (Barbanis \& Woltjer 1967), although the spiral supporting response of a disc to a massive object orbiting within it makes this distinction somewhat artificial (Julian \& Toomre 1966).  Subsequent development of these ideas (Lynden-Bell \& Kalnajs 1972; Carlberg \& Sellwood 1985; Binney \& Lacey 1988; Toomre \& Kalnajs 1991; Fuchs 2001) has shown that changes in the distribution of angular momentum among the stars should always be accompanied by an increase in random motion, or `heating', of the stellar distribution.  While the random velocities of stars have tended to increase over the lifetime of the Galactic disc, as noted above, even old disc stars today have $\Delta/R \ll 1$ and this fact indicates rather modest changes to the initial distribution of angular momentum.

A small change in the {\it distribution\/} of angular momentum does not, however, forbid large changes to the angular momenta of individual stars.  In this paper, we show that spiral waves in discs do alter stellar angular momenta while largely preserving the overall distribution of $L$ and causing little increase in random motion.  We argue that churning of the disc by spiral waves in this manner also moves clouds of interstellar gas, and we discuss observational support for, and some consequences of, this prediction.  The associated radial migration of stars and gas must reduce the metallicity gradient in galaxy discs and modify radically the distribution over metallicity of the stars that one now finds at a given radius.

As previously noted, the spiral response of a galactic disc to a co-orbiting mass clump blurs the distinction between scattering by spirals and by mass clumps.  In the present paper, we consider the collective response of the disc to such mass clumps (Toomre \& Kalnajs 1991) as adding to the transient spirals.  The direct scattering of individual stars by clouds is important, since it probably determines the shape of the local velocity ellipsoid (Carlberg 1987; Jenkins \& Binney 1990), but we show in the Appendix that the angular-momentum changes induced by star-cloud scattering are negligibly small.

We begin by summarizing the theory of scattering of stars by spiral waves.  We next present simulations which demonstrate that angular momentum exchanges at corotation, which have previously received little attention, are in fact the principal driver of radial migration in galaxies.

\section{Scattering by waves}

In this section we obtain relations between the angular momentum changes induced by spirals and the associated changes in the random velocities of stars that will be useful in Section 3, where we analyze numerical simulations.  

In the rotating frame of a steady spiral perturbation there is an `energy' invariant, Jacobi's integral $E_J$, which is defined as [Binney \& Tremaine (1987; hereafter BT) eq.\ (3-88)]
\be E_J=E-\Omega_pL,
 \ee where $E$ and $L$ are the specific energy and $z$-angular momentum of the star in a non-rotating frame and $\Omega_p$ is the pattern speed of the non-axisymmetric perturbation.  Hence, changes in energy and angular momentum are related by
\be\label{dEOdL}
\Delta E=\Omega_p\Delta L.
\ee

We now determine the fraction of the energy increment, $\Delta E$, which appears as random motion; the remainder changes the energy associated with circular motion.  If $J_R$ is any parameter that quantifies radial kinetic energy, we can obviously write
\be\label{chainr}
\d E={\pa E\over\pa J_R}\d J_R+{\pa E\over\pa L}\d L.
\ee
 If $J_R$ is chosen to be the `radial action', the partial derivatives in this expression become the angular frequencies, $\omega_R$ and $\Omega$, of a star's radial and azimuthal motion [BT eq.~(3-150a)].  In the epicycle approximation, $\omega_R=\kappa$ the classical epicycle frequency, $J_R$ is simply $E_R/\kappa$, and the radial kinetic energy $E_R$ is related to the maximum radial speed around an epicycle by $E_R=\fracj12(v_{R,{\rm max}})^2.$  Eliminating $\Delta E$ between equations (\ref{dEOdL}) and (\ref{chainr}), we obtain
 \be\label{DEDL}
\Delta J_R={\Omega_p-\Omega\over\omega_R}\Delta L.
\ee
 While changes in $L$ at corotation ($\Omega=\Omega_p$) do not cause changes in $J_R$, those away from corotation do.

\begin{figure}
\centerline{\psfig{file=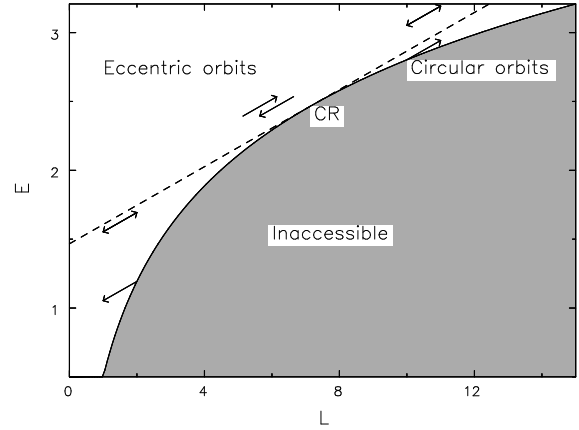,width=.9\hsize}}
\caption{Classical Lindblad diagram.  See text for a description. 
\label{Lindbladfig}}
\end{figure}

The classical Lindblad diagram, Fig.~\ref{Lindbladfig}, illustrates the physical origin of this relation.  Possible orbits occupy the upper left half of this diagram, and are bounded by the locus of the circular orbit of angular momentum $L$.  Equation (\ref{dEOdL}) states that a steady spiral wave moves stars on lines with slope $\Omega_p$.  Since the tangent to the locus of circular orbits at corotation has this slope, $J_R$, which is a measure of the distance from the circular-orbit curve, does not change to first order when a star corotating with the wave is scattered.  Scattering at more general locations, does result in energy being exchanged between random and orbital motion, and stars on near circular orbits must be preferentially scattered into non-circular orbits.

When a star is in Lindblad resonance with an $m$-armed spiral, its frequencies satisfy [e.g.\ BT eq.~(3-122)]
 \be\label{Lindblad}
\omega_R=\pm m(\Omega_p-\Omega),
\ee
 where the upper sign applies at the outer Lindblad resonance (OLR) and the lower sign applies at the inner resonance (ILR).  Combining equations (\ref{DEDL}) and (\ref{Lindblad}) we obtain the pleasingly simple result
 \be\label{DJDLwave}
\Delta J_R=\mp{1\over m}\Delta L\quad\hbox{(Lindblad resonance)}.
\ee
 Here the upper sign applies if the star lies at ILR ($\Omega>\Omega_p$) and the lower sign applies at OLR.  This relation is exact for nearly circular orbits, and remains the dominant term for moderately eccentric orbits where higher resonances also contribute (Lynden-Bell \& Kalnajs 1972).

\begin{figure*}
\centerline{\psfig{file=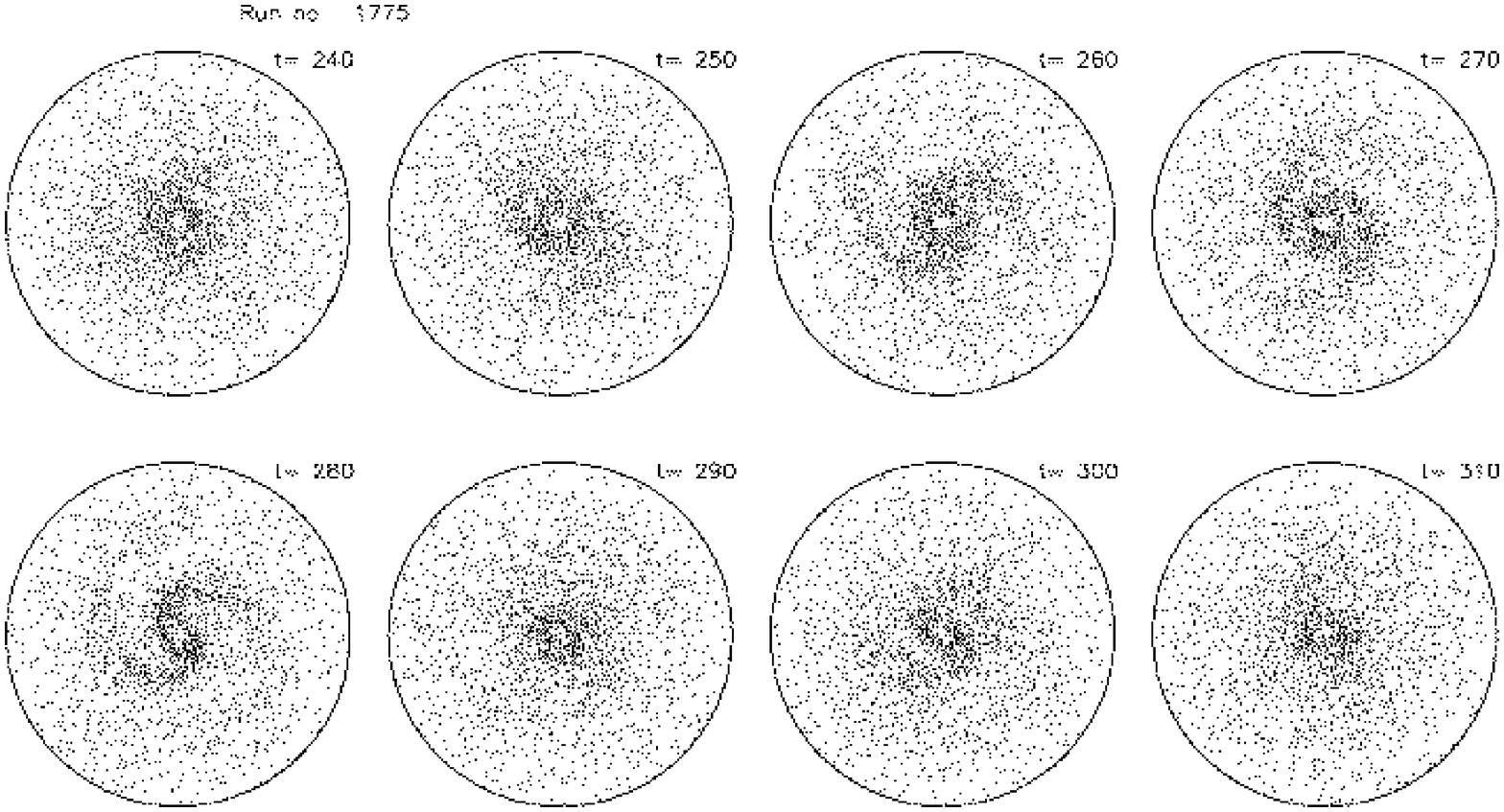,width=.9\hsize,clip=}}
\caption{The later part of the growth and subsequent decay of an isolated spiral mode in a disc that was seeded with a groove.  The radius of the circles is $15R_\i$ and only one particle in 120 is plotted. 
\label{singlespiral}}
\end{figure*}

\section{Single spiral}
As a test of the spiral wave theory presented in \S2, we begin by studying what happens in a simulation having a single transient spiral wave.  We later describe more realistic simulations with a succession of spiral waves.

\subsection{Choice of model}

We wish to construct a simulation that supports a single spiral wave standing out clearly from the noise; to achieve this goal, we follow the procedure described by Sellwood \& Kahn (1991, \S3.3).  Mestel's (1963) disc, in which the circular speed is everywhere $V_0$, suffers only from global one-armed instabilities when the disc is almost fully self-gravitating (Zang 1976; Toomre 1977, 1981; Evans \& Read 1998), and becomes stable to {\it all\/} global modes when only half the disc mass is active (Toomre 1981).  These remarkable stability properties have been confirmed in $N$-body simulations (Sellwood \& Kahn 1991; Sellwood \& Evans 2001).  An appropriate distribution function (DF) for this disc is (Zang 1976; Toomre 1977)
 \be
f_0(E,L) = {1\over2}{C L^q e^{[-E(q+1)]} \over [ 1 + (L_\i / L)^4] [ 1 + 
(L / 
L_{\rm o})^6 ]}. \label{df0}
\ee
 The constant $C$ [given in BT (eq 4-162)] yields the correct surface density when $f_0$ is integrated over all velocities with the factors in the denominator set to unity, while the pre-factor halves the active mass.  We set the parameter $q=11.44$, which in an infinite half-mass disc would yield $Q=1.5$.  The radial extent of the otherwise infinite disc is limited by the inner and outer tapers, represented by the factors in the denominator; these tapers are gentle enough not to provoke instabilities on their own (Toomre 1981).  The value of $Q$ is raised where the tapers kick in, since they reduce the active mass density by more than the velocity dispersion.  As the tapers also introduce length scales into the disc, we adopt the central radius of the inner taper, $R_\i = L_\i / V_0$, as our length unit, and set $L_{\rm o} = 15L_\i$.  We also adopt $V_0$ as our velocity unit, so $R_\i/V_0$ is our unit of time and $L_\i$ our unit of angular momentum.

Sellwood \& Kahn (1991) showed that the introduction of a narrow axisymmetric feature into the equilibrium DF destabilizes the disc in a predictable manner.  We therefore multiply $f_0$ by a function of $L$ and employ the distribution function
 \be
f(E,L) = f_0(E,L) \left[ 1 + {\beta w_{\rm L}^2 \over (L-L_*)^2 + w_{\rm L}^2}
\right].
\ee
 We chose the parameters of the Lorentzian function to be: $L_*=6.5L_\i$, $\beta=-0.4$, $w_{\rm L}=0.2L_\i$.  While this represents a groove in the angular momentum density, it causes only a mild depression in the surface density over a broad range of radii because the particles move on epicycles; the mean epicycle radius $\Delta \simeq 0.28R$.

\subsection{Numerical procedure}
We set up the model with considerable care to ensure an accurate initial equilibrium, and to keep random fluctuations to a minimum.  We adopt the quiet start procedure described by Sellwood \& Athanassoula (1986) to select the initial coordinates and place the particles at equal angular intervals around rings.

\begin{table}
\caption{Numerical parameters used in the two simulations}
\label{params}
\begin{tabular}{@{}lrr}
                  & Model S          & Model U \\
\hline
Grid size         & $230 \times 256$ & $110 \times 128$ \\
Softening length  & $0.05R_\i$       & $0.1R_\i$ \\
Active components & $m=2$ only       & $0 \leq m \leq 4$ \\
Particle number   & $6\times10^5$    & $10^6$ \\
Time step         & $0.005 R_\i/V_0$ & $0.005 R_\i/V_0$ \\
\hline
\end{tabular}
\end{table}

We use a highly-efficient PM method (Sellwood 1981, 1983) to compute the inter-particle gravitational forces, with a small softening parameter.  Particles move freely over a two-dimensional polar lattice of $(N_R,N_\phi)$ points, which is used to tabulate the gravitational field.  The grid points are spaced at $R=0.1R_\i[\exp(\alpha n) - 1]$, where $\alpha = 2\pi/N_\phi$ and $0 \leq n \leq N_R$.  The mass of each particle is spread over the four nearest grid points and we employ a standard Plummer softening formula to further smooth the forces at short range.  Radial and azimuthal acceleration components are computed directly and tabulated, avoiding the need to difference a potential function, and we use bi-linear interpolation to evaluate the acceleration components on each particle.  Our choices of numerical parameters in this run, model S, are given in Table~\ref{params}.

Forces from the particles are evaluated every time step.  Particles in the radial range $0.5<R<1$ are integrated using the given time step, while beyond $R=1$ time steps increase by factors of 2 for every factor of 2 increase in radius, with forces arising from particles in outer zones being determined by interpolation in time when needed to advance the motion of those on shorter steps -- see Sellwood (1985).  Time steps for the few particles inside $R=0.5$ are sub-divided by factors of 2 as they approach the centre, but the force field in this region, which is anyway dominated by the fixed central attraction, is not updated any more frequently.

The groove will excite a separate instability for each azimuthal wavenumber, $m$ (Sellwood \& Kahn 1991).  We therefore restrict the disturbance forces from the particles to those arising from the $m=2$ component of the density distribution.  The axisymmetric part of the field is held fixed throughout.

\begin{figure}
\centerline{\psfig{file=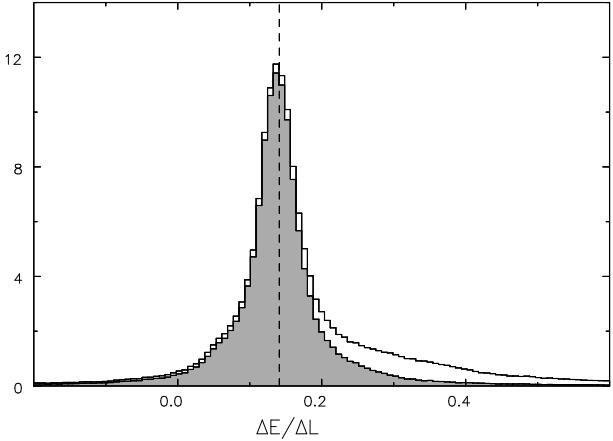,width=\hsize}}
 \caption{The distribution of the ratio $\Delta E/\Delta L$ of the changes in integrals up to $t=300$ in the disc that was seeded with a groove.  The vertical dashed line marks the measured pattern speed of the spiral mode.  The shaded histogram shows only those particles with initial $L>2$, while the unshaded shows all particles.
\label{Lorentzfig}}
\end{figure}

\subsection{Effect of one spiral disturbance}

\begin{figure*}
\centerline{\psfig{file=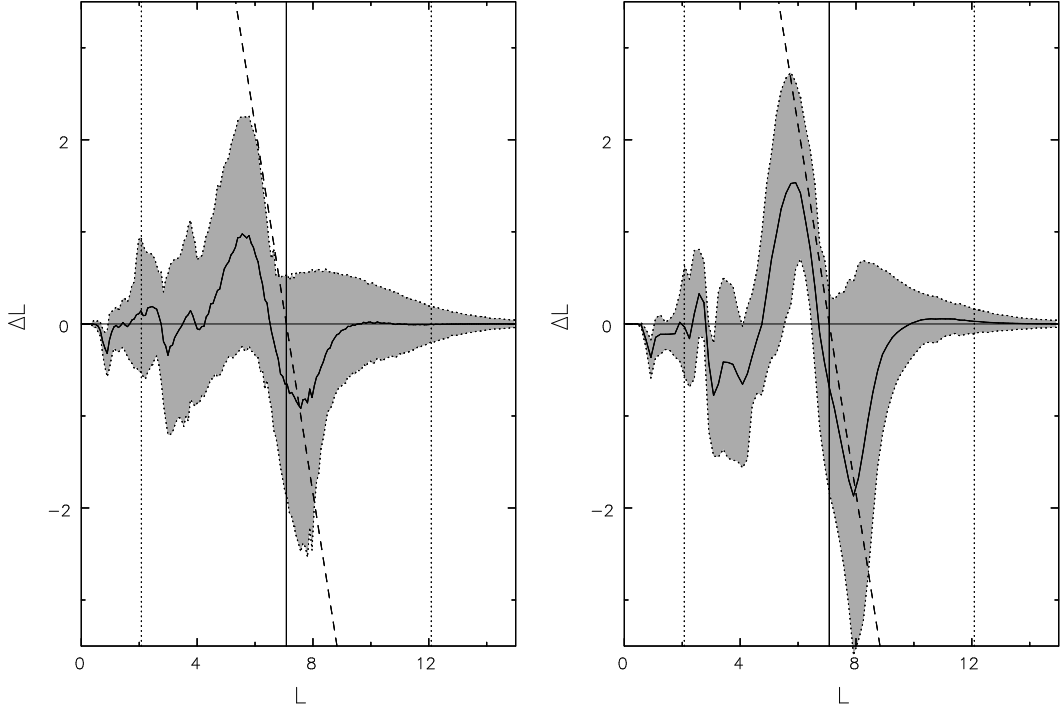,width=.8\hsize,angle=270}}
 \caption{Left panel: the full curve shows the mean change in $L$ to $t=300$ for all particles of the given initial $L$.  The shaded area is bounded by the 20th and 80th percentiles of the particles when ordered by magnitude of $\Delta L$.  The dashed line has slope $-2$, to show the locus of particles whose changes would be symmetric about corotation.  The vertical lines mark the positions of corotation (solid) and the Lindblad resonances (dotted) for near-circular orbits.
\label{Lchanges}}
\end{figure*}

Fig.\ \ref{singlespiral} shows the later evolution of a single growing spiral mode which saturates at about time 270 and then decays leaving a weak bar-like feature.  Fig.~\ref{Lorentzfig} shows the distribution of values of the ratio $\Delta E/\Delta L$, which eq.~(\ref{dEOdL}) implies will equal $\Omega_p$.  The sharp peak in this distribution coincides extremely well with the measured pattern speed of the mode, $\Omega_p=0.141$, implying that the vast majority of changes are induced by this spiral.  The shaded histogram shows this quantity evaluated for only those particles with initial $L>2L_\i$; a heavy tail to large values appears only when particles with small $L$ are included (unshaded), showing that the tail is associated with the formation of the inner bar.

Fig.\ \ref{Lchanges} shows the mean and the 20th and 80th percentiles of the distribution of $\Delta L$ values as a function of the initial $L$; corotation and the Lindblad resonances for the spiral are marked.  The left-hand panel shows the changes for all particles, the right-hand panel shows those for the 20\% of particles with the smallest epicycle energies.  The changes are large, especially near corotation.  $\Delta L$ increases with distance from corotation at such a rate that particles are transferred to the other side of corotation: the line of slope $-2$, on which particles move symmetrically across corotation, is marked.  Thus the net effect of the large individual changes in $L$ is to cause particles to change places.

Changes away from corotation are clearly much smaller.  Some are associated with the Lindblad resonances, but those near the centre are perhaps somewhat confused by the development of the bar.

\begin{figure}
\centerline{\psfig{file=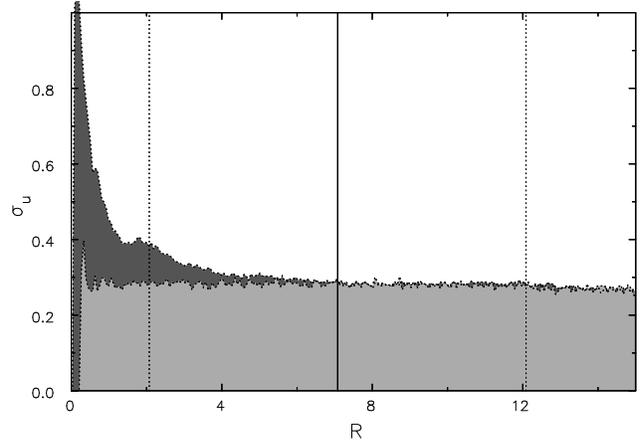,width=\hsize,clip=}}
\caption{Dispersion in radial velocity as a function of radius in the disc that was seeded with a groove at $t=0$ (light) and $t=300$ (dark).  The vertical lines mark the positions of corotation (solid) and the Lindblad resonances (dotted) for near-circular orbits.
\label{sigmau}} 
\end{figure}

The theory in Section 2 asserts that the heating associated with a given change $\Delta L$ tends to zero as we approach corotation from either side.  Fig.\ \ref{sigmau} confirms this prediction by showing the dispersion in the radial velocities as a function of radius at the initial moment and at $t=300$.  It is evident that the large changes in $L$ at corotation produced no significant heating whereas the comparatively small changes at the inner Lindblad resonance caused considerable heating.

\subsection{Orbit trapping at corotation}\label{trapping}

\begin{figure}
\centerline{\psfig{file=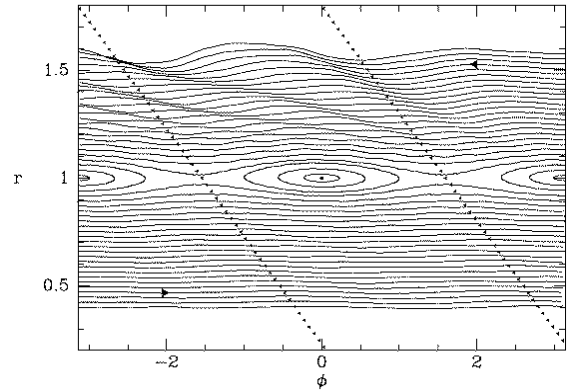,width=.9\hsize}}
 \caption{Orbits in a spiral perturbation. The background model is Mestel's disc and the perturbing potential is $\Phi_1=\Psi_0\cos[2(\phi-\phi_0-\Omega_pt)]$ where $\phi_0(R)=2(R/R_0-1)+\pi/2$.  The locus of the deepest part of the wave's potential is marked by dots.
\label{horse}} 
\end{figure}

The physics of the changes at corotation is easy to understand.  We show, in Fig.~\ref{horse}, orbits in a Mestel disc when a steady two-armed spiral perturbation is imposed.  Particles initially have velocity of magnitude $V_0$ in the tangential direction and are followed until their orbits approximately close on themselves.  Those inside corotation overtake the wave; they gain angular momentum as they fall into the spiral arm which causes them to move to larger radii and slow their drift relative to the wave, enabling large changes in $L$ to develop.  As the figure shows, particles initially close enough to the radius of corotation, interact so strongly that their speed relative to the wave is reversed.  These are called `horseshoe orbits' (Goldreich \& Tremaine 1982).  A particle that has moved out across corotation in this way subsequently slips backwards relative to the wave and eventually falls backwards into the other arm.  As it falls, it loses angular momentum, moves to smaller radii and eventually pulls ahead of the wave.  These successive episodes of angular-momentum gain and loss cause particles on horseshoe orbits to move around the maximum of the spiral potential as shown.

Section 3.3.3(b) of BT describes an analytic approximation to such motion in a sinusoidally varying potential of time-independent amplitude $\Psi_0$.  In this approximation, the phase variable $\psi=2(\phi-\phi_{\rm max})$, where $\phi_{\rm max}$ is the azimuth at which the potential peaks, obeys the pendulum equation
 \be\label{horseeqm}
\ddot\psi=-p^2\sin\psi,
\ee
 where
\be
p={2\over R}\sqrt{{|\Psi_0|A\over -B}}
\ee
 With $A$ and $B$ the Oort constants; for the Mestel disc, $A = -B = \Omega/2$.  Eq.~(\ref{horseeqm}) admits an energy invariant $E_p=\fracj12\dot\psi^2-p^2\cos\psi$.  Orbits with $E_p>p^2$ circulate, while the horseshoe orbits have $E_P<p^2$ and librate.  The periods of the horseshoe orbits increase with $E_p$ from the minimum value [BT eq.~(3-127b)]
 \be\label{horseP}
T_{\rm min}=\pi R_0\sqrt{{-B\over A|\Psi_0|}}
\ee
 that is associated with the orbits that make the smallest excursions in $R$.  For the orbit on which $R$ varies most, the changes in $R$ and $L$ are given by [BT eq.~(3-129)]
 \begin{eqnarray}\label{horseL}
\Delta R&=&2\sqrt{{|\Psi_0|\over-AB}}\nonumber\\
\Delta L&=&2AR\Delta R.
\end{eqnarray}

\begin{figure}
\centerline{\psfig{file=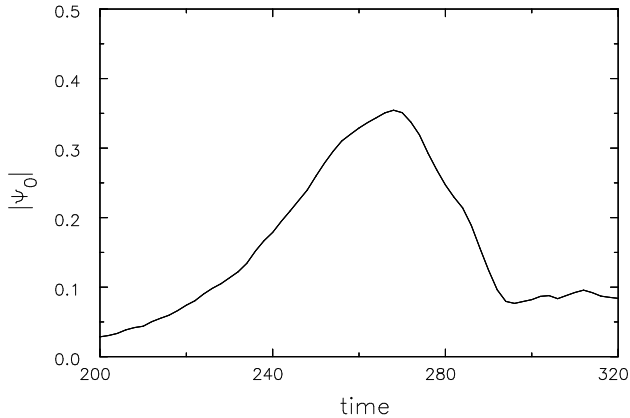,width=\hsize}}
 \caption{Part of the time variation of the amplitude of the $m=2$ component of the gravitational potential, $|\Psi_0|$, measured at $R=7.1R_\i$, the radius of corotation of the spiral shown in Fig.~\ref{singlespiral}.
\label{Psi0}} 
\end{figure}

The spiral perturbation in the simulation is transient, rather than steady.  Clearly, scattering from one side of corotation to the other cannot be completed if the non-axisymmetric potential is at full strength for less long than $0.5T_{\rm min}$.  On the other hand, if the disturbance persists for a time much greater than $0.5T_{\rm min}$, many particles will be scattered both across and back, leading to smaller net changes.

The time variation of $\Psi_0$ near the radius of corotation is shown in Figure \ref{Psi0}.  From $t=0$ to $t \sim 250$, the disturbance potential grows exponentially, with growth rate $\gamma = 0.047$.  At $t\sim250$ growth slows and from $t\sim270$ the disturbance potential fades even more quickly than it grew.  Rough estimates from Fig.~\ref{Psi0} are that the peak amplitude of the perturbing potential $\Psi_0 \simeq 0.35$ and its duration $T_{\rm on} \sim 20$ time units.  With eq.~(\ref{horseP}) these values yield $T_{\rm min} \sim 38$, so $T_{\rm on} \sim 0.53 T_{\rm min}$.  Hence, as expected, particles cross the resonance at most once.  Inserting our value for $\Psi_0$ in eq.~(\ref{horseL}) we find that the largest change in angular momentum is predicted to be $\Delta L = \Delta R \simeq 2.4$, in excellent agreement with the observed magnitude of the changes around corotation in Fig.~\ref{Lchanges}.

This picture of the instability's physics can only be approximate since it is based on orbits in a steady rather than a time-varying potential.  None the less, we have shown that it yields a semi-quantitative understanding of the simulations. 

Moreover, it predicts that particles on nearly circular orbits will experience larger changes in $L$ than particles on more eccentric orbits: the angular velocity with which the latter circulate about the galactic centre varies significantly as the particles oscillate radially, so it is impossible for such particles to hold station with respect to a steadily rotating wave.  The right-hand panel of Fig.~\ref{Lchanges} confirms this prediction by showing the mean and spread of changes in $L$ for the 20\% of particles with the smallest epicycle energies.  One sees that the changes are $\sim50\%$ larger than those for the disc as a whole.

The emergence of horseshoe behaviour may be the principal change that limits the amplitude of a spiral mode.  Standard linear stability theory (e.g.\ Kalnajs 1971, 1977) assumes an infinitesimal disturbance and computes the orbital response as small departures from the unperturbed orbits.  The linear mode is the self-consistent disturbance for which the sum of the orbit responses is equal to that required to generate the disturbance potential with a particular pattern speed, $\Omega_p$, and growth rate, $\gamma$.  At any finite amplitude there will be particles on horseshoe orbits, but the existence of these orbits will not have a significant impact at first, because (a) they will be small in number, and (b) they will have long periods, and first-order perturbation theory will give a good account of the motion of a particle on a horseshoe orbit until the particle reaches the point at which $\dot\psi=0$.  As the mode grows the period of any trapped orbits decreases, and the width of the horseshoe region grows exponentially; thus the number of particles that develop horseshoe behaviour grows exponentially and the time for which linear perturbation theory remains valid for them becomes short.  When the condition $\gamma T_{\rm min}>1$ is first violated, particles on horseshoe orbits that have passed through $\dot\psi=0$ are suddenly important contributors to the overall dynamics, and we have to expect linear perturbation theory to fail.  Thus many particles reverse their direction of motion in the rotating frame rather abruptly, which leads to a rapid decrease in the disturbance potential as the density enhancement disperses on a timescale $\ga 0.5T_{\rm min} \; (\simeq 20)$.  These ideas are consistent with the behaviour shown in Fig.~\ref{Psi0}.

\begin{figure*}
\centerline{\psfig{file=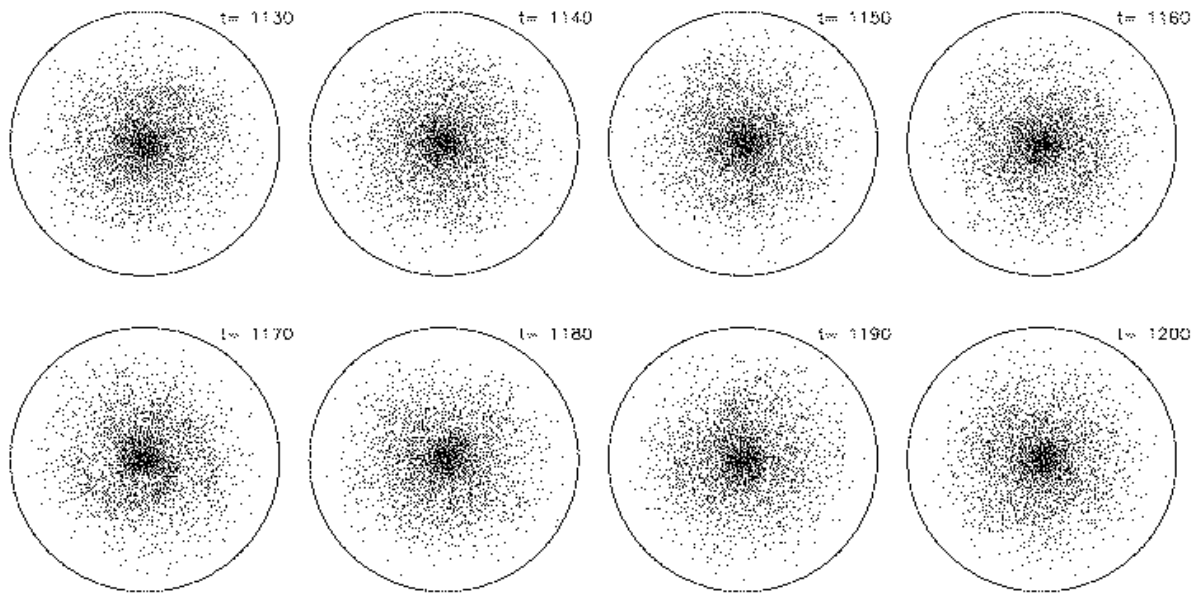,width=.9\hsize,clip=}}
 \caption{Part of the evolution of the unconstrained simulation.  The spiral activity is very mild at all times, the pattern visible at time 1170 is perhaps the strongest of all.  The radius of the circles is $21R_\i$ and only 1 particle in 200 is plotted.
\label{unconst}}
\end{figure*}

\subsection{Comparison with other work}
Lynden-Bell \& Kalnajs (1972; hereafter LBK) presented an analysis of the angular-momentum exchanges between a growing wave and a disc that went far beyond the elementary considerations of Section 2.  Their widely quoted result is that in the limit of vanishing growth rate angular-momentum changes occur only at resonances.  Carlberg \& Sellwood (1985; hereafter CS), focusing on heating, extended their analysis to transient waves.

LBK averaged the change in $L$ over all initial phases of the star.  Clearly, orbit-averaging makes it impossible to discover how wide is the spread in $\Delta L$ at a given initial value of $L$.  However, one might hope that it would correctly predict the net change in $L$, and qualitatively it does; their results for growing waves correctly predict that resonances are broadened, and that stars just inside corotation should, on average, gain $L$ while there should be a net loss just outside.

Since the second-order perturbation theory developed by LBK and CS assumes that changes to all quantities are small, their analyses implicitly exclude the trapping process associated with horseshoe orbits (Fig.~\ref{horse}), but may give an adequate description of the behaviour outside the trapping region.  The signs of the angular momentum changes in Fig.~\ref{Lchanges} are in agreement with LBK and CS, but the largest changes in $L$ are associated with order unity changes to an angle variable and are therefore outside the regime where their predictions hold.  Furthermore, any perturbation analysis which seeks to expand changes in the dynamical variables in powers of the perturbing potential $\Psi_0$, will have difficulty in recovering the result $\Delta L\sim\sqrt{|\Psi_0|}$.  Horseshoe orbits become important in the simulation both because the perturbation is strong and because it is of short duration.  A stong perturbation causes the horseshoe region to have a significant width, while a more enduring perturbation would allow particles to recross corotation and end up with essentially unchanged angular momenta.

In an influential paper Wielen (1977) showed that the increase in the velocity dispersion of K and M stars with age could be reproduced by a model in which stars diffused in velocity space with diffusion coefficients that are isotropic and constant.  In reality, even in the absence of scattering, the velocity of a star changes radically in the course of a half an epicycle period, so a picture based on velocity space is of limited value.  When stars are viewed as diffusing in integral space (Binney \& Lacey 1988), one finds both that diffusion is extremely anisotropic, and that the diffusion coefficients vary from point to point in the space in a manner which contradicts Wielen's simple picture.  Indeed, star-wave scattering drives diffusion that is as anisotropic as it logically can be, in that stars diffuse along lines in two-dimensional integral space (Section 2).  Neglecting such effects leads to seriously mistaken results.

It seems possible that the horseshoe behaviour we have identified is the effect that Zhang (1996) describes as a ``collisionless shock'' as stars pass through spiral arms.  She also claims additional scattering in this event, but any scattering at corotation must occur without heating the disc.

\section{Unconstrained simulation}

We next describe a more realistic simulation which displays a succession of very mild spiral waves.  We again use a half-mass Mestel disc with initial $Q=1.5$ (in the absence of tapers).  This model has no initial groove and we place our one million particles at random azimuths, so that the model starts with shot noise.  We also include forces from azimuthal wavenumbers $0 \leq m \leq 4$, so that there are often significant patterns with different rotational symmetries present at the same time.  The numerical details, summarized under Model U in Table~\ref{params}, differ in insignificant respects from those in Model S described above.

\begin{figure}
\centerline{\psfig{file=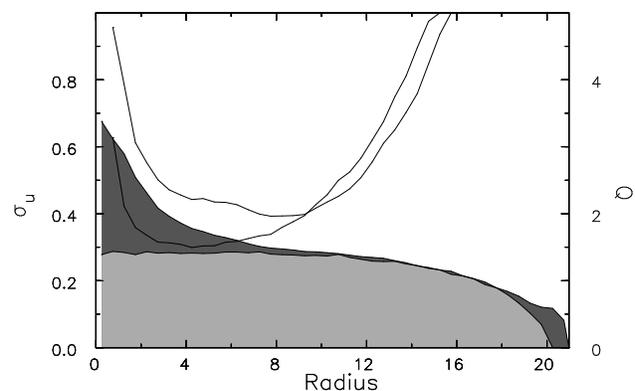,width=\hsize}}
 \caption{The light shaded region shows the initial radial velocity dispersion in model U, the dark shaded region shows the change by $t=1200$.  The two curves and right-hand scale show the local stability parameter $Q$ at the same times.
\label{Qplot}}
\end{figure}

The simulation, shown in Fig.~\ref{unconst}, displays a succession of very mild spiral waves that gradually heat the inner disc, while the outer disc has not heated much by $t=1200$.  The change in the radial velocity dispersion between the start and the end of the simulation is illustrated in Fig.~\ref{Qplot}; it more than doubles for stars in the inner disc.  The stability parameter, $Q$, rises to $Q>2$ everywhere by $t=1200$, when spiral activity has diminished but not completely ceased.

\begin{figure}
\centerline{\psfig{file=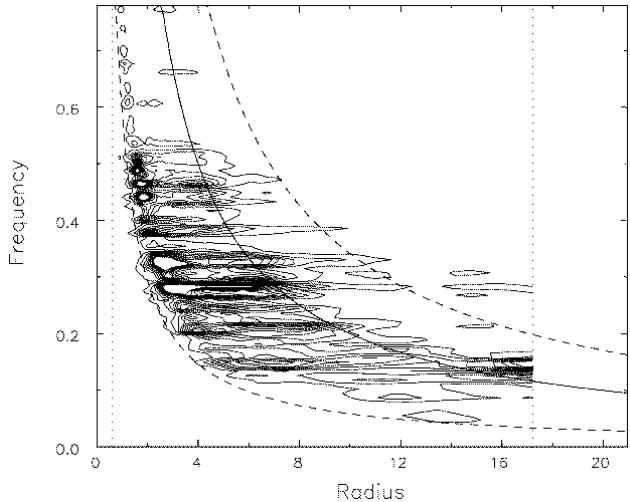,width=\hsize}}
 \caption{The power spectrum of the $m=2$ component of the density variations in the unconstrained simulation.  The dashed curves mark the loci of ILR and OLR for the given frequency, while the full curve shows the locus of CR.
\label{powfig}}
\end{figure}

Fig.~\ref{powfig} shows as a function of radius the power in the Fourier transform with respect to time of the $m=2$ component of the density distribution.  The power is strongly concentrated around a number of discrete frequencies $\omega_i$.  The Lindblad resonances for each frequency occur where the horizontal ridge intersects the dashed curves and the intersection with the full curve marks corotation.  It is evident that there is always significant power extending from the ILR to CR for each wave, and the power sometimes extends out to the OLR.

\begin{figure}
\centerline{\psfig{file=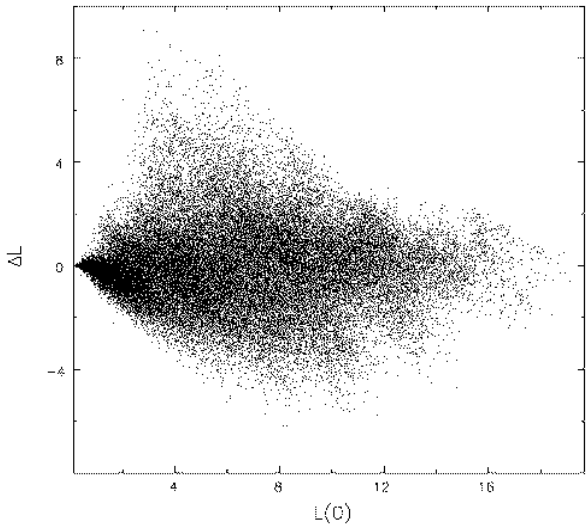,width=\hsize,clip=0}}
 \caption{The change in $L$ to $t=1200$ plotted against the initial $L$ for one particle in 11 of those in the simulation with recurrent transient spiral structure.
\label{messyfig}}
\end{figure}

\begin{figure}
\centerline{\psfig{file=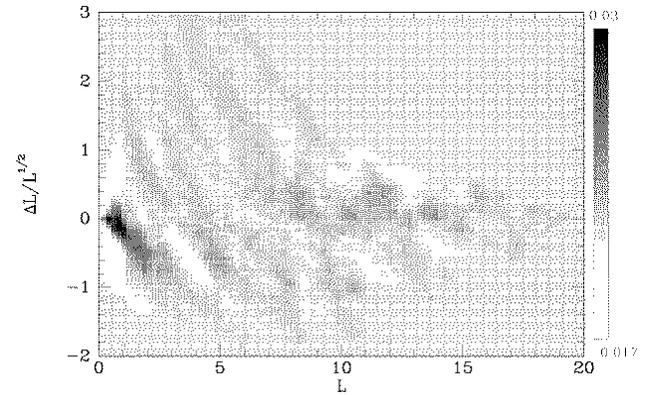,width=\hsize}}
 \caption{An unsharp-masked version of Fig.~\ref{messyfig} reveals ridges in the particle distribution that extend diagonally from top left to bottom right.\label{stripefig}} 
\end{figure}

Fig.~\ref{messyfig} shows for each particle the difference, $\Delta L$, between the initial and final values of $L$ for that particle plotted against the initial value of $L$.  Changes in $L$ of 50 percent are not uncommon.  Fig.~\ref{stripefig}, which is an unsharp-masked version of Fig.~\ref{messyfig}, reveals small-scale variations in particle density that in Fig.~\ref{messyfig} are obscured by the overall decline in the particle density from the line $\Delta L = 0$.  Several diagonal ridges of enhanced particle density are apparent.  Each ridge is presumably the signature of a spiral feature of some frequency $\omega_i$, and is equivalent to the shaded strip in Fig.~\ref{Lchanges} as it runs from upper left to lower right through CR.  Thus each diagonal ridge in Fig.~\ref{stripefig} is created by particles on horseshoe orbits crossing corotation for some $\omega_i$.

Figure \ref{rchanges} shows the extent of mixing within the disc that spiral activity causes by showing several distributions in final home radius of particles that initially have essentially identical home radii.  The distributions are wide: particles at intermediate radii can be moved by a factor 2 or more in either direction.  The initial mean epicycle size, $\Delta \simeq 0.28R_{\rm home}$ throughout most of the disc, doubles for stars in the inner disc, while those at large radii are scarcely changed by $t=1200$ (Fig.~\ref{Qplot}).

\section{Discussion}
\subsection{Stellar migration}

\begin{figure}
\centerline{\psfig{file=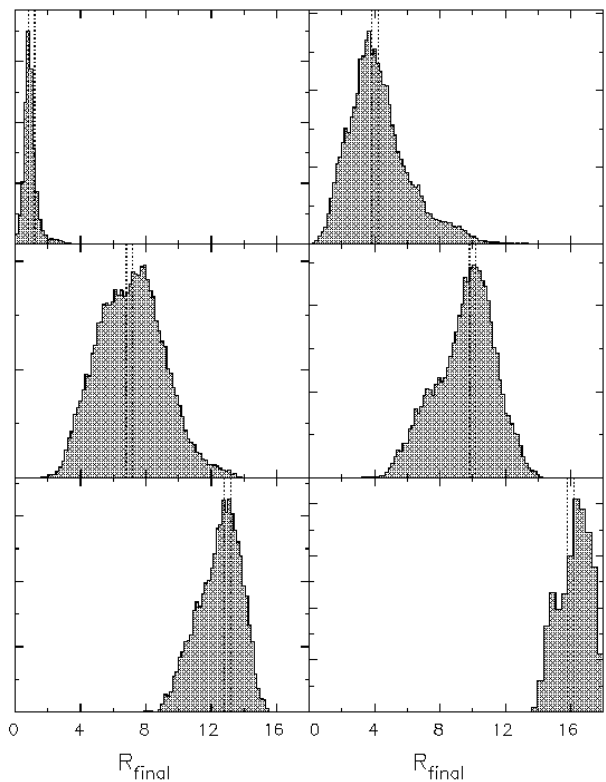,width=\hsize,clip=}}
 \caption{The distributions of home radii at $t=1200$ of particles starting with initial home radii within the range indicated by the vertical dotted lines.  Typical epicycle radii are initially $\sim 0.28R_{\rm home}$; they do not increase much in the outer disc but rise in the inner disc to $\sim 0.5R_{\rm home}$ by $t=1200$.
\label{rchanges}}
\end{figure}

Our simulations suggest that stars can migrate significant distances within discs, whilst remaining on nearly circular orbits.  To decide how important this effect is for real galaxies, we need to scale the numerical results to take into account that galaxies are typically three times older than the duration of our simulation, and to allow for possible differences in the amplitude $|\Psi_0|$ of the non-axisymmetric components of the potentials.  In Section 3.4, we argued that the size of a typical step in radius is proportional to $\sqrt{|\Psi_0|}$, so, if successive steps are uncorrelated and on average equally spaced in time, the net distance migrated should scale like $\sqrt{t|\Psi_0|}$, where $t$ is the age of the disc.

Near-infrared photometry provides the obvious way to estimate $\Psi_0$.  Rix \& Rieke (1993) have argued that the contribution by massive stars to $K$-band luminosity is small and patchy, so the $K$-band light is a reliable measure of the stellar mass distribution in galaxies on large scales.  Rix \& Zaritsky (1995), Gnedin et al.\ (1995), Block \& Puerari (1999) find that in the $K$-band the arm-interarm surface-brightness contrast is generally $\ga30\%$.  For comparison in our unconstrained simulation the arm-interarm contrast in mass density is at most $\sim15\%$.  The extent of radial migration must also be a sensitive function of the disc-mass fraction, because both the arm-interarm contrast and the magnitude of the potential fluctuation that is produced by a given contrast increase with disc mass-fraction.  A variety of arguments indicate that in high surface-brightness disc galaxies, more than half of the mass interior to two scale lengths lies in the disc (Debattista \& Sellwood 1998, 2000; Weiner, Sellwood \& Williams 2000; Kranz, Slyz \& Rix 2002; Binney \& Evans 2001).  In our simulations exactly half of the mass resides in the disc, so we are probably underestimating the extent of radial migration by a factor of a few. Widening the distribution shown in the left-hand middle panel of Fig.~\ref{rchanges} by even a factor $1.5$, one is led to the conclusion that old stars formed in the solar neighbourhood should be scattered nearly uniformly within the annulus from $R=4\kpc$ to $R=12\kpc$.

If the radial migrations of individual stars were equivalent to the random steps taken by particles in a diffusive medium, the galactic disc would spread quite rapidly because the radial steps are rather large.  In the present case however, any spreading of the disc will actually be rather slow because the large migrations that occur each side of corotation almost exactly cancel by conservation of angular momentum; only the relatively small transfer of angular momentum outwards from the ILR to corotation enables the disc to spread radially.  The inability of angular-momentum exchanges at corotation to induce radial spreading is reflected in their inability to heat the disc significantly [eq.~(\ref{DEDL})], since when a disc spreads, random energy is inevitably released (LBK; Lynden-Bell \& Pringle, 1974).

\begin{figure}
\centerline{\psfig{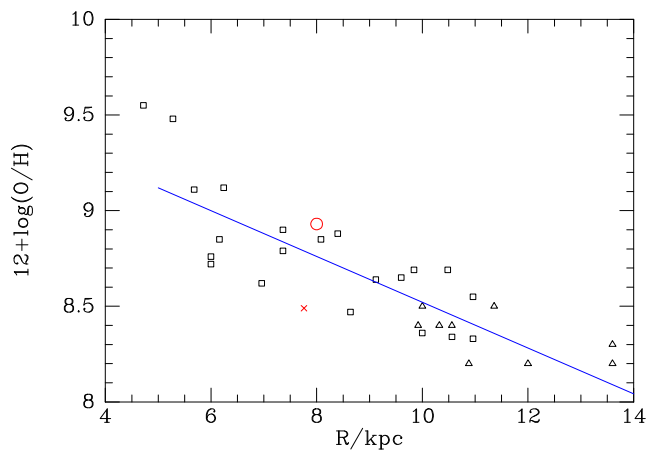}}
 \caption{Oxygen abundances of 31 HII regions from Shaver et al.\ (1983; squares) and from Vilchez \& Esteban (1996; triangles).  The original data have been rescaled to $R_0=8\kpc$.  The Orion nebula, from Shaver et al., is marked by a cross, and the Sun, with O abundance 8.93 from Anders \& Grevesse (1989), is marked by a circle.  The line is the linear least-squares fit to the data for $R>5.5\kpc$.
\label{Shavfig}}
\end{figure}

\subsection{Abundance distributions}
The really important effects of radial migration lie in the field of chemical evolution.  Fig.~\ref{Shavfig} plots measured values of O/H against Galactocentric radius for HII regions from Shaver et al.\ (1983) and Vilchez \& Estaban (1996) rescaled to $R_0=8\kpc$.  The line is the linear least-squares fit to the data for $R>5.5\kpc$.  The dispersion in [O/H] about the mean line is $0.16\,$dex, which is not significantly larger than the quoted errors in the data of Shaver et al.\ and Vilchez \& Estaban.  Hence, it is widely believed that the intrinsic dispersion in [O/H] at fixed $R$ is $\la0.1\,$dex (Edmunds 1998) and at each radius in the disc we may suppose that the ISM has a well-defined metallicity.

\begin{figure}
\centerline{\psfig{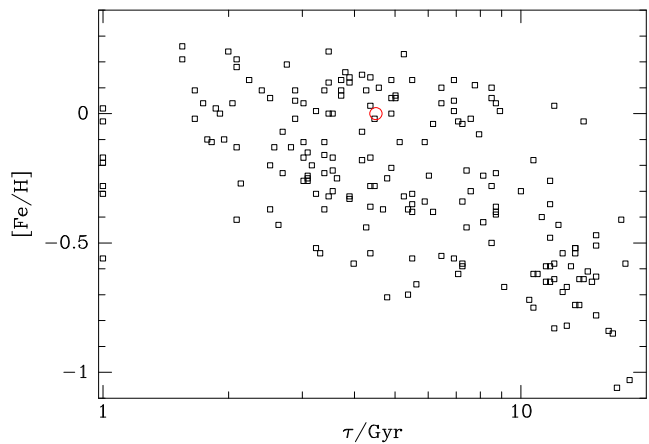}}
 \caption{The distribution of a sample of 189 nearby stars in metallicity and age.  The data are from Edvardsson et al.\ (1993).  The circle marks the position of the Sun.
\label{Edfig}}
\end{figure}

\begin{figure}
\centerline{\psfig{file=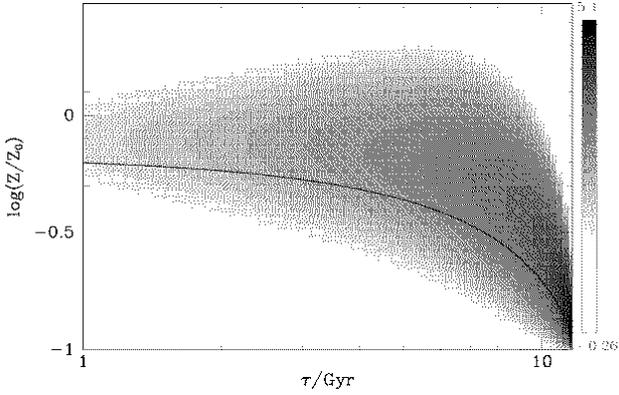,width=\hsize}}
 \caption{The density of points in the $(\log(Z),\log\tau)$ plane predicted by the models of chemical evolution and stellar migration described in the text. The curve shows the local metallicity of the ISM according to equations (\ref{Zfirst}) and (\ref{Zsecond}).
\label{nztau}} 
\end{figure}

The squares in Fig.~\ref{Edfig} show the distribution in [Fe/H] and age of 189 stars from the sample of Edvardsson et al.\ (1993).  The Sun is plotted as a circle.  Since Edvardsson et al.\ selected stars to obtain wide ranges in $\tau$ and [Fe/H], the distribution of points has significant selection bias in the sense that the spread in age and [Fe/H] is unrepresentatively broad (see \S3.2 of Edmunds 1998).  None the less, the figure shows that age and metallicity are weakly correlated if at all; the only clear effect is that there is a dearth of young, metal-poor stars, presumably because the ISM at all star-forming radii has been for some time more metal-rich than $\hbox{[Fe/H]}\sim-0.4$.

Radial migration enables old stars that formed from promptly-enriched gas at small galactocentric radii to appear in a solar-neighbourhood sample, thus weakening any correlation between age and metallicity.  Migration from outside $R_0$ also allows a smaller number of young metal-poor stars to be in such a sample.  A simple quantitative model illustrates these phenomena.  In this simple model, at time $t$ and radius $R$ stars form at a constant rate from an ISM having a metallicity rising linearly with time as
 \be\label{Zfirst}
Z(R,t)=Z_{\rm G}+{t\over t_0}[Z(R,t_0)-Z_{\rm G}],
\ee
 where $Z_{\rm G}$ is the metallicity of protodisc material and $t_0=12\Gyr$ is the age of the disc.  At the present time the metallicity of the gas is given by the line in Fig.~\ref{Shavfig}
 \be\label{Zsecond}
\log[Z(R,t_0)]=a-b\left({R\over R_0}-1\right),\hbox{ where }
\left\{\matrix{a=-0.17\cr b=0.96.}\right.
\ee
 The density of stars in the $(\log Z,\log\tau)$ plane is then given by
 \be
N_{R_0}(\tau,Z)\propto
\tau ZR\e^{-R/R_\d}P_\tau(R_0|R)\left|{\d R\over\d Z}\right|.
\ee
 Here the surface density of newly formed stars is assumed to be proportional to $\e^{-R/R_\d}$ with $R_\d=0.28 R_0$ (Drimmel \& Spergel, 2001), while $P_\tau(R_0|R)$ is the probability that a star formed at radius $R$ migrates to $R_0$ in time $\tau$, and $R(t_0-\tau,Z)$ is the radius at which the ISM had metallicity $Z$ at time $t_0-\tau$.  We calculate illustrative predictions for the case in which $P_\tau$ is
 \be
P_\tau(R_0,R)=(2\pi\sigma_\tau^2)^{-1/2}
\exp\left[-{(R_0-R)^2\over2\sigma_\tau^2}\right]
\ee
with
 \be
\sigma_\tau=R_0\left[(0.16)^2+(0.4)^2{\tau\over t_0}\right]^{1/2}.
\ee
 The first term in the square brackets represents the radial migration associated with epicyclic motion for a typical thin-disc radial velocity $\sigma_u=34\kms$ (Table 10.4 of Binney \& Merrifield 1998), which we assume for simplicity to be time-independent.  Fig.~\ref{nztau} shows the predicted distribution that one obtains from these formulae with $Z_{\rm G}=0.1Z_\odot$, which is the approximate lower boundary on the populated zone in Fig.~\ref{Edfig}. The distribution of Fig.~\ref{nztau} has features in common with the observational plot, Fig.~\ref{Edfig}. In particular both plots show a dearth of stars around $(2,-0.5)$, a high population density around $(8,0)$ and an approximately horizontal upper boundary on the populated zone.  The observational plot does not show the same extremely high density of stars around $(10,-0.75)$ that is evident in the predicted distribution.  This density peak is large due to our assumption that at early times the ISM everywhere had metallicity $0.1Z_\odot$, but it is exacerbated by the use of a logarithmic age coordinate. In the observational plot errors in age estimates, which are obviously largest for the oldest stars, scatter stars from the furthest realistic ages to unrealistically high ages, thus reducing the density of stars around $\tau=t_0$.

Wielen et al.\ (1996, hereafter WFD) argue that the Sun is a beautiful example of stellar migration.  It is on an unusually circular orbit for a star of its age ($4.5\Gyr$), with home radius about $200\pc$ outside $R_0$ and pericentre about $140\pc$ inside $R_0$.  WFD note that the Sun is not only more metal-rich than the average nearby star of comparable age by $0.17\,$dex (Fig.~\ref{Edfig}), but is actually more metal-rich than the present-day local ISM (as represented by the Orion molecular cloud, which is marked by a cross in Fig.~\ref{Shavfig}).  They argue from these facts that the Sun was born $R\sim(R_0-2\kpc)$.

Fig.~\ref{nztau} clearly shows that migration brings into the solar neighbourhood many more stars that are more metal rich than the local ISM than it imports more metal-poor stars: metal-rich immigrants are imported by arms with corotation inside $R_0$ that effect exchanges between the solar neighbourhood and an annulus around $R\sim5\kpc$, while metal-poor immigrants are imported by arms that have corotation outside $R_0$ and move stars from $R\sim11\kpc$.  Since there are many fewer stars at $R\sim11\kpc$ than at $R\sim5\kpc$, metal-rich immigrants outnumber metal-poor ones by a significant factor.  The clear implication is that at $R\sim5\kpc$ there should be many stars with sub-solar metallicities.  As most migration is accomplished with little change to the random motion of the stars, it is not surprising that the Sun has a nearly circular orbit.

\subsection{Interstellar gas}

Gas clouds are deflected by the spiral-arm potential in much the same way as are stars.  The deflections generate in-going and out-going streams that occur at different azimuths and do not cross until large changes in radius have been achieved (Sellwood \& Preto 2002).  The resulting flow in the corotation annulus has a characteristic anticyclonic form.  Our model predicts that the entire flow lasts no more than half its turnover time and will involve velocities $|v_r| \la |\Psi_0|/(-BR_0)$.  Fridman, et al.\ (2001) have already called attention to flows of this form in NGC 3631, and derive peak radial velocities approaching half the orbital speed at their estimated corotation radius, again suggesting a very strong spiral pattern.

Fig.~\ref{Lchanges} showed that stars with lower random velocities experienced larger changes in $L$ than did hotter stars.  Since gas clouds have smaller random velocities than any population of stars, they should experience still larger changes in $L$.  Hence, it is possible that the Orion nebula lies so far below the mean line in Fig.~\ref{Shavfig} because it has immigrated to the solar neighbourhood from $R\sim10\kpc$, where its metallicity would be typical.  If the metallicity distribution of clouds at a given radius has non-negligible width, the theoretical stellar metallicity distribution of Fig.~\ref{nztau}, which is based on the assumption of negligible width, will be too narrow at small ages. Comparison of Figs~\ref{Edfig} and \ref{nztau} suggest that this may be true, but one has to bear in mind that the width of the observational distribution is exaggerated by both selection bias and errors in the measured stellar metallicities.

\subsection{Dust grains}
The impact of migration on the metallicity of interstellar gas must be to some extent masked by mixing following collisions between clouds of different metallicity, to produce a new cloud of intermediate metallicity.  The chemical composition of an individual interstellar grain cannot change, however, so the compositions of the interstellar grains that are currently located at a given radius should show similar evidence for migration to that evident in Fig.~\ref{Edfig}.  Clayton (1997) presents evidence for just such an effect.

As the disc ages, the isotopes $^{29}$Si and $^{30}$Si, which are secondary, should become more abundant relative to $^{28}$Si, which is primary.  Dust grains that were formed before the solar system have been identified in meteorites, and measurements of the relative abundances of the three Si isotopes (Hoppe et al.\ 1994a,b) confirm the theoretical prediction (Clayton 1988; Timmes \& Clayton 1996) that the abundance ratios $n(^{29}\hbox{Si})/n(^{28}\hbox{Si})$ and $n(^{30}\hbox{Si})/n(^{28}\hbox{Si})$ should be tightly correlated in a positive sense.  Three problems mar the beauty of the observed correlation: the slope of the correlation differs from that expected; the Sun's ratios are barely compatible with the correlation; and the measured ratios indicate that the majority of presolar grains formed from material that was substantially more heavily processed, nucleosynthetically, than that for which the Sun formed.  Clayton (1997) points out that this last finding is similar to the metal-richness of the Sun relative to the local ISM because it implies that material that left the ISM at some time prior to the formation of the Sun, somehow experienced more nucleosynthesis than material that stayed in the ISM, and continued to be enriched, right up to the formation of the Sun.  Clayton further argues that the large changes in home radius proposed by WFD could explain the Si isotope ratios in presolar grains by making it credible that they formed from material that had been enriched by stars that themselves formed $\sim2\kpc$ interior to the birthplace of the Sun. In the light of our results we suggest that these grains themselves migrated from $R\sim4\kpc$ to $R\sim6\kpc$, where they were incorporated into the protosolar cloud.

\subsection{Galactic dynamos}
Radial mixing in the ISM may also help with the well-known problem posed by the large-scale (ordered) component of {\bf B}-fields in galaxies (e.g.\ Rees 1994).  Standard $\alpha\Omega$-dynamo theory (Parker 1955) is thought to yield too low a growth rate to achieve the present-day observed field strengths (Beck, et al.\ 1996) from the likely seed fields.  The growth-rate is proportional to the geometric mean of the rates of galactic shear (the $\Omega$ term) and cyclonic circulation (the $\alpha$ term) (Kulsrud 1999).  Current estimates of the $\alpha$-term are based on supernovae-driven turbulence (Ferri\`ere 1998), but the large-scale radial mixing discussed here may enhance the $\alpha$-effect substantially, and thereby increase the growth-rate obtainable from the dynamo.

\section{Conclusions}
We have shown that the dominant effect of spiral waves is to churn the disc.  Stars just inside corotation swap places with those outside because each group surfs on opposite sides of the wave.  We present a simplified model of this behaviour based on the theory of horseshoe orbits in a steady potential.  We find that transient spiral arms are at peak strength for long enough to produce generally only a single crossing of corotation by a large fraction of the stars over a broad swath of the disc around this resonance.  The length of the largest steps taken in $R$ scales as $\sqrt{|\Psi_0|}$, where $\Psi_0$ is the peak amplitude of the perturbing potential.  Scaling to real galaxies requires calibration of the near IR photometry and estimates of the disc mass fraction, but step sizes of 2 -- 3 kpc are expected.

Groups of stars move in both directions, and generally just exchange places.  This exchange occurs without any significant radial spreading of the disc or increase in non-circular motions. In this respect exchanges at corotation differ significantly from those at the Lindblad resonances, which do engender heating and spreading.  We emphasize, therefore, that the dominant scattering process in the disc does not alter the surface density profile of the disc.

We argue that the peak amplitude and duration of a spiral wave are limited by horseshoe behaviour at corotation.  The density enhancement disperses rapidly when a substantial fraction of particles on horseshoe orbits have reversed their motion in the rotating frame.

The induced radial migration of stars will largely erase the correlation between metallicity and stellar age that is a clear prediction of standard chemical-evolution theory.  It will also introduce into the solar neighbourhood large numbers of stars that are substantially more metal-rich than the local ISM was at the time of their birth.  Both effects are in accord with data for the solar neighbourhood.  Finally, radial migration of stars will gradually flatten any radial metallicity gradient within the disc.  We predict that there should be many stars with sub-solar metallicities at $R\sim 5\kpc$.  The fact that metallicity gradients survive in disc galaxies in the teeth of such mixing puts a stronger requirement on the effectiveness of whatever process is responsible for their creation.

Deflection of interstellar clouds by spiral arms at corotation should induce an anticyclonic circulation in the ISM that may already have been detected in observations of interstellar velocity fields.  The circulation will bring into the solar neighbourhood gas from both low- and high-metallicity regions.  Such importation of non-standard gas may account for the anomalously low metallicity of the Orion molecular cloud, and for the presence in the protosolar cloud of dust grains with large ratios $n(^{29}\hbox{Si})/n(^{28}\hbox{Si})$ and $n(^{30}\hbox{Si})/n(^{28}\hbox{Si})$, characteristic of highly nucleosynthetically processed material.

Our picture is incomplete, however, since we have implicitly assumed that the spiral behaviour manifested by the simulations correctly mimics the phenomenon in galaxies.  Since the origin of spirals in galaxies, and in the simulations themselves, continues to lack a convincing explanation (e.g.\ Sellwood 2000), this logical gap is likely to persist for some time.  We regard the ability of simulations to predict the age-velocity dispersion relation of local disc stars (Carlberg \& Sellwood 1985), the shape of the velocity ellipsoid (Jenkins \& Binney 1990), the need for a dissipative component to maintain persistent spiral patterns (Sellwood \& Carlberg 1984), and the present success in accounting for the spread of metallicity of older stars, as indirect evidence to suggest that the recurrent transient patterns in the simulations do indeed mimic the behaviour in real galaxies.

\section*{Acknowledgments}

We are indebted to S. D. Tremaine for invaluable comments as referee of this and earlier drafts of the paper.  This work was supported in part by NSF grant AST-0098282 to JAS.

\section*{Appendix: Star-cloud scattering}

Mass clumps on near-circular orbits induce a collective spiral response, or wake, from the surrounding disc (Julian \& Toomre 1966).  We regard such collective effects as part of the spiral structure of the disc which is the subject of the main part of the paper.  Here we simply show that, when the collective wake is ignored, the interactions between stars and the mass clump can be neglected as a possible source of additional angular momentum redistribution.

For small radial oscillations, there is a convenient expression for a star's epicycle energy in terms of the components $u_R$ and $u_\phi$ of its velocity with respect to the local standard of rest [BT eq.\ (7-94a)]:
 \be
E_R=\fracj12(u_R^2+\gamma^2u_\phi^2),
\ee
 where $\gamma=2\Omega/\kappa$; i.e.\ $\gamma = \sqrt{2}$ for a flat rotation curve.  Small oscillations of a star perpendicular to the plane approximately conserve the integral
 \be
E_z=\fracj12u_z^2+\Psi(R,z),
\ee
 where $\Psi$ is an effective gravitational potential.  Following Spitzer \& Schwarzschild (1953) we consider stars to be scattered in the impulse approximation, in which the scattering is complete in a distance small compared to both the epicycle amplitude and the smallest distance over which the Galactic potential changes significantly.  In this approximation we have
 \begin{eqnarray}\label{DeltaEE}
\Delta E_R&=&u_R\Delta u_R+\fracj12(\Delta u_R)^2\nonumber\\
&&\qquad+\gamma^2[u_\phi\Delta u_\phi+\fracj12(\Delta u_\phi)^2\big]\\
\Delta E_z&=&u_z\Delta u_z+\fracj12(\Delta u_z)^2.\nonumber
\end{eqnarray}

When a star encounters a massive object such as a GMC or a star cluster, the recoil of the object is negligible and it is natural to study the encounter in the frame of reference in which the scatterer is stationary throughout the encounter.  If, moreover, the scatterer is moving on a circular orbit, this preferred reference frame is a steadily rotating one, and Jacobi's integral may be written [BT eq.~(3-90)]
 \be
E_J=\fracj12|\b v|^2+\Phi_{\rm eff}(\b r),
\ee
 where $\b v$ is the star's velocity in the cloud's rotating frame and $\Phi_{\rm eff}$ is the sum of the gravitational and centrifugal potentials.  Since we are treating encounters in the impulse approximation, we may neglect the change in $\Phi_{\rm eff}$ between the in- and the out-states of the scattering event, and conclude from the invariance of $E_J$ that the scattering merely redirects the velocity of the star in the rotating frame without changing its magnitude $v$.  Moreover, at the radius of the cloud, $\b u=\b v$.  We have therefore
 \begin{eqnarray}
0&=&\Delta u^2=2u_R\Delta u_R+(\Delta u_R)^2\nonumber\\
&+&2u_\phi\Delta u_\phi+(\Delta u_\phi)^2+
2u_z\Delta u_z+(\Delta u_z)^2.
\end{eqnarray}
 Comparing this with equation (\ref{DeltaEE}) and noting that $\Delta L=R\Delta u_\phi$, we obtain
 \be\label{bigdE}
\Delta E_R+\Delta E_z=(\gamma^2-1)\left(u_\phi
+\fracj12\Delta u_\phi\right){\Delta L\over
R}.
\ee
If we define 
\be
J_{\rm rand}\equiv{1\over\Omega}\big(E_R+E_z),
\ee
then equation (\ref{bigdE}) can be written
\be\label{dJdLcloud}
\Delta J_{\rm rand}=(\gamma^2-1){u_\phi+\fracj12\Delta u_\phi
\over v_c}\Delta L,
\ee
 which is analogous to our result (\ref{DEDL}) for spiral waves.  When a cloud scatters a star of the same specific angular momentum, $u_\phi=0$ and the change in $J_{\rm rand}$ vanishes to first order because the interaction occurs at corotation for the star.  Heating occurs when a cloud scatters a star that has a different specific angular momentum because then $u_\phi\ne0$ and the interaction occurs away from corotation.

Since a small fraction of stars pass a cloud with $u_\phi\simeq0$, most scattering by clouds is associated with heating.  Hence we may use estimates of the amount of heating in discs to constrain the amount of angular momentum transfer for which clouds are responsible.  Equation (\ref{dJdLcloud}) implies that
 \be
{\Delta J_{\rm rand}\over\Delta L}\simeq {u_\phi\over v_c}.
\ee
 Comparing this result with equation (\ref{DJDLwave}) for $u_\phi\la50/\gamma \kms$, we see that star-cloud encounters are about three times more effective at changing $L$ for a given increment in random velocity than two-arm spirals at Lindblad resonance.  Jenkins \& Binney (1990) concluded that $\ex{(\Delta J_R)^2}_{\rm wave}\simeq90\ex{(\Delta J_R)^2}_{\rm cloud}$.  Here angle brackets imply averages over all encounters for a star that has $E_R/E_z\ga1$ and an epicycle radius that is not large compared with the typical radial wavenumber of spiral structure.  We have that
 \begin{eqnarray}
{1\over 90}\ex{(\Delta J_R)^2}_{\rm wave}&\simeq&
\ex{(\Delta J_R)^2}_{\rm cloud}
\sim \ex{(\Delta J_{\rm rand})^2}_{\rm cloud}\nonumber\\
&\simeq&{R^2(\gamma^2-1)^2\over v_c^2} \\
&\times&\ex{u_\phi^2(\Delta u_\phi)^2+u_\phi(\Delta u_\phi)^3 
+\fracj14(\Delta u_\phi)^4} _{\rm cloud}.\nonumber
\end{eqnarray}
 The ensemble average on the right hand side of this equation is dominated by the first term, since the second averages to something near zero and the third term cannot exceed $\fracj14$ of the first and is likely to be much smaller.  Since scattering events occur at all radial phases, we may use the approximation $\ex{u_\phi^2(\Delta u_\phi)^2}\simeq\ex{u_\phi^2}\ex{(\Delta u_\phi)^2}$.  With equation (\ref{DJDLwave}) and setting $\gamma=\sqrt2$, we have finally 
 \be
\ex{(\Delta L)^2}_{\rm cloud}\sim
{1\over90}{v_c^2\over m^2\ex{u_\phi^2}}\ex{(\Delta L)^2}_{\rm Lindblad}.
\ee
 Hence, $\ex{(\Delta L)^2}_{\rm cloud}=\ex{(\Delta L)^2}_{\rm Lindblad}$ when $\ex{u_\phi}^{1/2}=v_c/(9.5m)\simeq12\kms$ for $v_c=220\kms$ and $m=2$.  Since $12\kms$ is a fairly typical value of $\ex{u_\phi}^{1/2}$ for an old disc star, we conclude that the angular-momentum changes induced by clouds are comparable to those induced by waves at Lindblad resonance.  Since we have shown that waves induce much larger changes in $L$ near corotation, clouds are not significant drivers of radial migration overall.

\end{document}